# Fast and Accurate Transaction Level Modeling of an Extended AMBA2.0 Bus Architecture

Young-Taek Kim, Taehun Kim, Youngduk Kim, Chulho Shin, Eui-Young Chung,
Kyu-Myung Choi, Jeong-Taek Kong, Soo-Kwan Eo
CAE Center, SOC R&D, System LSI, Samsung Electronics Co., Ltd.
Giheung-Eup, Yongin-City, Gyeonggi-Do, Korea  449-711
ytkimn@samsung.com

## Abstract

*Transaction Level Modeling (TLM) approach is used to meet the simulation speed as well as cycle accuracy for large scale SoC performance analysis. We implemented a transaction-level model of a proprietary bus called AHB+ which supports an extended AMBA2.0 protocol. The AHB+ transaction-level model shows 353 times faster than pin-accurate RTL model while maintaining 97% of accuracy on average. We also present the development procedure of TLM of a bus architecture.*

## 1. Introduction

Platform based design has become a common methodology as the complexity of SoC design increases. The major advantage of this method is to maximally exploit the reusability of IP components, hence reduce the time-to-market pressure.

But, the one of main challenges in the platform based design is how to exploit the optional architecture, which requires highly abstracted simulation models to maximize simulation with reasonable accuracy.

Transaction level modeling (TLM) has been increasingly favored to address this issue.[1] This method requires more effort than in pure functional modeling, but generates more realistic traffics to communication architecture.

Communication architecture models also need to satisfy the same requirements for IP component models. Moreover, these models have to consider other requirements. First, the model should be flexible enough to consider various topology configurations. Second, it should be easy to model and evaluate several arbitration schemes. Third, interface should be clearly and easily modeled to integrate IP component models efficiently. Fourth, a good analysis environment should be tied with the model to assess the simulation results. For example, bus contention, utilization and throughput are very important metrics to be assessed in communication architecture.

## 2. AHB+ Bus System Architecture

AMBA2.0 protocol is widely being used, but the serious problem is that it cannot guarantee master's QoS (Quality of Service). AHB+ is designed to address this issue.[2] Also, it aims at improving the overall throughput as well. AHB+ maximizes bus utilization and guarantees master's QoS using several arbitration algorithms with request pipelining scheme.

AHB+ bus architecture consists of AHB+ main bus, DDR Controller (DDRC), and a Bus Interface (BI). BI is designed for transferring special information between arbiter and memory controller such as the next transaction information, idle bank, access permission and so on.

In order to guarantee QoS of IPs, AHB+ has special internal registers. These registers store QoS objective value and the type of real-time/Non-real time master. To maximize bus throughput, AHB+ hides the latencies incurred between the requests of masters by pipelining the master requests. In AHB+ bus architecture, the arbiter gives the next transaction information to DDRC in advance, then, DDRC can pre-charge the next accessed memory bank, thus it is possible to suppress the latency between transactions. As a result, the next data can be served immediately right after the previous data is processed. This is the concept of bank interleaving which maximizes bus utilization.

## 3. Transaction Level Modeling of AHB+ Bus System

We describe our transaction-level modeling steps of AHB+ and its memory controller as follows.

### 3.1. Re-definition of Protocol in Transaction-level

First of all, the AHB+ protocol needs to be redefined as transaction-level ports of TLM. Unfortunately, in many cases, bus protocol in a design specification is described at signal level. Therefore, it is necessary to map signals into TLM transaction-level ports. (typically, they are implemented as variables or functions.)

### 3.2. Behavior Description at Transaction Port

The next step is to model the behavior of each transaction-level port. For example, in RTL, a master can immediately get 'HGRANT (bus grant signal)' from the bus after sending 'HBUSREQ (bus request signal)'. This step is represented as the transaction port of a master calls *CheckGrant()* and receives 'true' as a return value. After then, the master sends 'HADDR (address)' and 'HRDATA (read data)' and receives 'HREADY'. This behavior is also modeled as the transaction port of the master calls '*Read(addr, *data, *ctrl)*' function and receives 'OK' as a return value.

### 3.3. Function Description in Bus Internals / Memory Controller

Internal functions of the arbiter were implemented. In the design of AHB+, seven arbitration filters are implemented and they are always activated without the consideration of master / slave combinations. In addition, we modeled the write buffer of AHB+ for the purpose of processing write transactions more speedy and efficiently. The write buffer stores the information of write transactions when a master cannot get a bus grant at the right time. The write buffer behaves as another master when it is occupied by waiting transactions.

We also modeled the DDR controller as a TLM in order to increase the accuracy of overall communication architecture. This is very important that the overall latency of data access from memory critically depends on it. To maximize memory bandwidth, each bank has a state machine separately, and column, row, and pre-charge accesses have different priorities by scheduling scheme. To increase the cycle accuracy, we modeled the FSM as accurate as register transfer level. Instead, the data path is highly abstracted to increase simulation speed.

### 3.4. Design of Special Interfaces and Platform Integration

AHB+ and DDRC are interfaced with a special protocol called BI (Bus Interface). This interface is designed to support the bank interleaving feature for throughput enhancement.

To consider the cycle accuracy of communications through the bus architecture, we defined the timings of each transaction function. After defining the interface signals, AHB+ bus main and DDR controller were connected each other. Then whole system was completed together master and memory.

### 3.5. Assertion for Error Handling

We have inserted two types of assertion statements into the transaction-level models. The first feature is for the functional debugging of the model itself. The other feature is related to the property checking and this is very helpful especially when the bus model is integrated with master models and simulated for performance analysis.

### 3.6. Insertion of Profiling Features

We also provide convenient profiling features by integrating the model with the commercial EDA tool. In AHB+ TLM, we implemented bus and master port profiling features in transaction-level ports and some internal functions such as arbiter, write buffer and so on.

### 3.7. Flexibility and Reusability

For the flexibility and reusability, AHB+ TLM has several parameters, such as bus width, write buffer depth, arbitration algorithm on/off, and etc. Other parameters are selection of real-time/non-real time type of a master, write buffer on/off, and QoS value.

## 4. Model Accuracy and Simulation speed

The transaction-level AHB+ model are validated by comparing it with the RTL model. To increase simulation speed, we used method-based modeling method rather than thread-based method. Also, we used 2-step cycle-based simulation tool to further speed up the simulation.

We modeled and simulated a target system by changing the traffic patterns of the maters as shown in Table 1. From Table 1, the average accuracy difference is below 3%. From this fact, our AHB+ model is accurate enough to be used for performance analysis.

The simulation speeds were measured at both TL and RTL. At RTL, it is 0.47 Kcycles/sec, and at TL, 166 Kcycles/sec. When we used only one master as an input for evaluating the pure performance of our bus architecture, the simulation speed went up to 456 Kcycles/sec.

The implemented bus TLM boosts simulation speed with sufficient accuracy and provides bus profiling features for performance evaluation. From the experimentation results, the implemented model is 353 times faster than RTL model while maintaining 97% of accuracy on average.

## References

[1] L. Cai, and D. Gajski, "Transaction level modeling : an overview", in *IEEE/ACM/IFIP International Conference on Hardware/Software Codesign and System Synthesis*, pp.19-24, 2003.
[2] K-M. Lim and Y-D. Kim, "Samsung DVDP Bus & Memory Architecture", Samsung, 21 May. 2002

**Tab. 1 : Simulation Results**

| | | RTL | TLM | Cycle Diff. | Accuracy Diff. (%) |
|---|---|---|---|---|---|
| Read (12 Read Master) | Single | 25505 | 25222 | 283 | 1.11 |
| | Burst4 | 25028 | 24264 | 764 | 3.05 |
| | Burst8 | 48398 | 48649 | −251 | −0.52 |
| | Mixed | 31380 | 33084 | −1704 | −5.43 |
| Write (12 Write Masters) | Single | 27268 | 24273 | 2995 | 10.98 |
| | Burst4 | 26188 | 24286 | 1902 | 7.26 |
| | Burst8 | 49455 | 48610 | 845 | 1.71 |
| | Mixed | 31262 | 28851 | 2411 | 7.71 |
| Read/Write (Read:8, Write:4) | Single | 27674 | 25767 | 1907 | 6.89 |
| | Burst4 | 25141 | 25263 | −122 | −0.49 |
| | Burst8 | 50463 | 51465 | −1002 | −1.99 |
| | Mixed | 33747 | 34384 | −637 | −1.89 |
| Total | | 401509 | 394118 | 7391 | 28.41 |
| Average | | 33459 | 32843 | 616 | 2.37 |